\providecommand\options{dvips,letterpaper,aps,prl,amsmath}
\documentclass[\options,preprint]{revtex4}
\usepackage{amssymb,amsthm}
\usepackage{xcolor,graphicx}
\usepackage{hyperref}

\renewcommand*{\Re}{\mathop{\mathrm{Re}}\nolimits}

\newcommand*{\abs}[2][]
{#1\lvert{#2}#1\rvert}
\newcommand*{\nm}[2][]
{#1\lVert{#2}#1\rVert}
\newcommand*{\exv}[2][]
{#1\langle{#2}#1\rangle}
\newcommand*{\bra}[2][]
{#1\langle{#2}#1\rvert}
\newcommand*{\ket}[2][]
{#1\lvert{#2}#1\rangle}
\newcommand*{\braket}[3][]
{#1\langle{#2}#1\vert{#3}#1\rangle}
\newcommand*{\bracket}[4][]
{#1\langle{#2}#1\lvert{#3}#1\rvert{#4}#1\rangle}

\newcommand*{\f}[2]{\ifcase#1%
	\mathit{#2}\or\mathrm{#2}\or%
	\mathcal{#2}\or\mathbb{#2}\or%
	\mathfrak{#2}\or\mathbf{#2}\or%
	\mathsf{#2}\fi%
}

\makeatletter
\def\@bibdataout@aps{%
 \immediate\write\@bibdataout{%
  @CONTROL{%
   apsrev41Control%
   \longbibliography@sw{%
    ,author="00",editor="1",pages="1",title="0",year="0"%
   }{%
    ,author="48",editor="1",pages="0",title="",year="1"%
   }%
  }%
 }%
 \if@filesw
  \immediate\write\@auxout{\string\citation{apsrev41Control}}%
 \fi
}%
\makeatother

\DeclareMathOperator{\tr}{tr}

\newcommand*{\maintitle}%
{Standard Quantum Limit and Heisenberg Limit in Function Estimation}
\newcommand*{\UTPhys}%
{Department of Physics, University of Tokyo, 7--3--1 Hongo, %
 Bunkyou-ku, Tokyo, 113--0033, Japan}
\newcommand*{\RIKEN}%
{RIKEN Center for Emergent Matter Science (CEMS), Wako, Saitama, 351--0198, Japan}
\newcommand*{\InsPI}%
{Institute for Physics of Intelligence, University of Tokyo, 7--3--1 Hongo, %
 Bunkyou-ku, Tokyo, 113--0033, Japan}

\begin{document}
\title{\maintitle}
\date{\today}
\author{Naoto Kura}\affiliation{\UTPhys}
\author{Masahito Ueda}\affiliation{\UTPhys}\affiliation{\RIKEN}\affiliation{\InsPI}
\begin{abstract}
 Unlike well-established parameter estimation, function estimation faces conceptual and mathematical difficulties despite its enormous potential utility. 
 We establish the fundamental error bounds on function estimation in quantum metrology for a spatially varying phase operator, where various degrees of smooth functions are considered.
 The error bounds are identified in both cases of absence and presence of interparticle entanglement, which correspond to the standard quantum limit and the Heisenberg limit, respectively.
 Notably, these error bounds can be reached by either position-localized states or wavenumber-localized ones.
 In fact, we show that these error bounds are theoretically optimal for any type of probe states, indicating that quantum metrology on functions is also subject to the Nyquist-Shannon sampling theorem, even if classical detection is replaced by quantum measurement.  
\end{abstract}
 \maketitle
 Accurate estimation of signals with a limited amount of resource is a fundamental problem in physics.
 Quantum metrology has made a profound contribution to this problem by demonstrating a classically unattainable scaling of the estimation error~%
 \cite{Holland1993,Giovannetti2004,Giovannetti2006,Higgins2007,Giovannetti2011}.
 This non-classical accuracy, called the \emph{Heisenberg limit}, can be reached by various forms of quantum features including entanglement~\cite{Giovannetti2004}, quantum circuit~\cite{Higgins2007}, bosonic~\cite{Anisimov2010,Gagatsos2016} and spin~\cite{Kitagawa1993,Wineland1992} squeezing, and quantum statistics~\cite{Datta2012}.
Applications of the Heisenberg-limited measurement range from detection of fundamental signals such as atomic clocks~\cite{DeBurgh2005,Kessler2014}, gravitational waves~\cite{Schnabel2010,LIGOCollab2011}, a scalable cat state~\cite{Gao2010} and polariton condensates~\cite{Deng2007,Huang2012}.

 The Heisenberg limit is significantly better than the \emph{standard quantum limit} (SQL) $\delta = O(N^{-1/2})$, which sets the accuracy bound arising from uncorrelated noises, where $N$ is the size of resource.
 While the Heisenberg limit $\delta = O(N^{-1})$ and the SQL $\delta = O(N^{-1/2})$ apply to both scalar estimation and vector estimation~\cite{Macchiavello2003,Ballester2004,Humphreys2013,Yao2014,Szczykulska2016,Tsang2017}, it is generally considered that the continuity of the signal may alter the scaling law.
 Such a problem can be categorized as function estimation, which has attracted growing attention.
 For example, atomic clocks~\cite{Galleani2008,Fraas2016} and graviational waves~\cite{LIGOCollab2011,Kolkowitz2016} involves time-varying singals, which offer richer information when treated as functions.
 More generally, exploration for new phenomena involves observing structures in a continuous in space and/or time, which can be represented as functions.
 This indicates that functional structures play crucial roles in generic continuous systems including measurements on magnetometry~\cite{Budker2007,Vengalattore2007}, nanostructured materials~\cite{Johnson1998,Tsang2009,Brida2010,Samantaray2017}, live cells~\cite{Stephens2003,Michalet2005}, and event horizons~\cite{EHTCollab2019b,EHTCollab2019a}.
 Hence, it is not only a fundamental question but also relevant to a wide range of applications to ask how quantum metrology can contribute to the detection of functions with ultimate accuracy.

 The quantum version of function estimation has been investigated in terms of the signal detection theory in Refs.~\cite{Berry2002,Berry2006,Tsang2011,Berry2015,Fraas2016,Dinani2017}.  In fact, weaker scaling laws are implied when the target parameter can change continuously in time, such as a Gaussian signal.
 The demonstration of such unconventional limits has recently become within the experimental reach due to the realization of, e.g., high-N00N states~\cite{Afek2010} and optical phase tracking~\cite{Yonezawa2012}.
 Although the detection theory is applicable to stochastic noises, it does not support the case where the relevant parameter is not inherently stochastic, which is often the case with quantum imaging and quantum signal processing~\cite{Mamaev2003,Galleani2008,Brida2010,Lupo2016}.

 In this Letter, we present a fundamental framework of quantum metrology on functions.
 Unlike parametric estimation, function estimation involves infinite degrees of freedom and inevitably requires further assumptions on the target function.
 Assuming only the smoothness of the function, we find the SQL of $O(N^{-q/(2q+1)})$ and the Heisenberg limit of $O(N^{-q/(q+1)})$, where $q$ indicates the degree of smoothness of the function.
 Our framework allows analysis of estimation errors of data series under given smoothness, such as a bound on the amplitude of derivatives.
 This includes the previous results on Gaussian processes through computation of their smoothness~\cite{Belayev1961}, as demonstrated later.
 The data series requires neither to have a prior distribution nor even to be continuous, allowing, for example, a sample with a finite number of discontinuous points~\cite{Brida2010,Samantaray2017}.
 Moreover, we have found that the error limit can equally be saturated by states which are localized in position or wavenumber.
 This result implies the equivalence between space discretization and momentum cutoff in quantum information processing, reminiscent of the Nyquist--Shannon sampling theorem in classical statistics.

 \paragraph{Setup.}
 \begin{figure}[bth]
  \includegraphics[width=0.7\columnwidth]{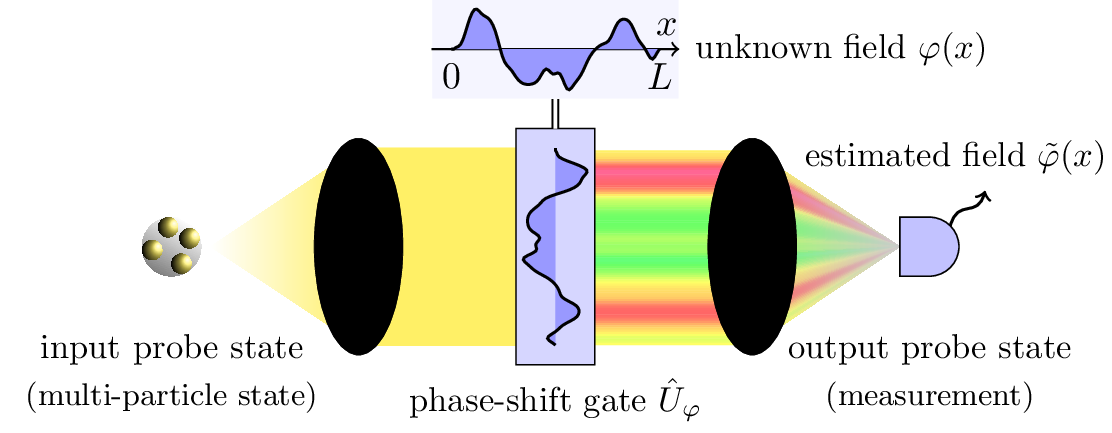}
  \caption{\label{Figure-FQM}%
  (Color online) Schematic illustration of quantum estimation of functions.
  First, some multi-particle state is prepared as an input probe state.  It then passes through a phase-shift gate $\hat{U}_{\varphi}$ generated by a spatially varying field $\varphi(x)$ that we want to know.  Finally, the output probe state is measured, whence the estimated function $\tilde{\varphi}(x)$ is computed.
  }
 \end{figure}
 We consider the estimation of an unknown function $\varphi(x)$ defined over the interval $0\le x\le L$ by using the position-dependent phase-shift gate $\hat{U}_\varphi$ shown in Fig.~\ref{Figure-FQM}.
 For simplicity, we also assume the periodic boundary condition.

 To estimate $\varphi$, we prepare an $N$-particle state as a probe, which evolves according to the unitary operator $\hat{U}_{\varphi}$ and then is measured.
 These particles are distributed in the interval $[0,L]$ and have two internal states: one state $\ket{+}$ interacts with the phase-shift gate $\hat{U}_{\varphi}$, and the other state $\ket{-}$ does not.

 We work in the first-quantization formalism and denote by $\ket{x;\pm}$, which is the position eigenstate at $x$ with internal state $\pm$.
 Let the phase-shift gate act as
 \begin{math}
  \hat{U}_{\varphi} \ket{x; +} = e^{i\varphi(x)} \ket{x; +}
 \end{math}
 and
 \begin{math}
  \hat{U}_{\varphi} \ket{x; -} = \ket{x; -}.
 \end{math}
 The unknown function $\varphi$ can be estimated from measurement on this output.
 When the probe is composed of $N$ separable particles, the error of the function estimation is bounded by the SQL.
 A probe with appropriately entangled $N$ particles, on the other hand, leads to the Heisenberg limit.

 We recall that the estimation error of a scalar parameter $\theta$ is computed as $\delta^2 = \f3E[\abs{\tilde\theta-\theta}^2]$, where $\tilde{\theta}$ is the estimator depending on the stochastic nature of measurement outcome.
 Similarly, we consider a stochastic estimator $\tilde{\varphi}$ for the function, and compute the mean-square periodic error (MSPE) ~\cite{Routtenberg2013} as
 \begin{equation}
  \delta^2 = \f3E\biggl[ \int_0^L \frac{dx}{L} [\tilde{\varphi}(x)-\varphi(x)]_{2\pi}^2 \biggr].
   \label{def-MSPE}
 \end{equation}
 In other words, the estimation error is averaged over $x$ and the modulus is replaced by $[\tilde{\varphi}(x)-\varphi(x)]_{2\pi} = \min_{n\in\f3Z} \abs[\big]{\tilde{\varphi}(x)-\varphi(x)+2\pi n}$, i.e.\@ the minimal modulus regarding the $2\pi$ periodicity of a phase.

 The main difficulty in function estimation lies in the fact that the problem involves an the infinite degrees of freedom.
 In particular, the lower bound on $\delta^2$ cannot be established for an arbitrary function, since we cannot exclude any rapidly fluctuating functions from a finite number of measurements.
 Hence, we impose the following constraint on the target function $\varphi$:
 \begin{equation}
  \int_0^L \frac{dx}{L} \abs[\big]{\varphi'(x)}^2 \le \frac{M^2}{L^{2}}
   \label{DiffReg}
 \end{equation}
 for some positive number $M>0$.
 With this constraint, we can establish a suitable lower bound on sufficiently smooth and slowly varying functions $\varphi$.

 The condition \eqref{DiffReg} can be applied only when the target function is differentiable.
 In our framework, we consider more general functions without differentiability: the H\"older continuity $\abs{\varphi(x+\epsilon)-\varphi(x)} = O(\epsilon^q)$ for fixed $0<q\le 1$~\cite{Lunardi2012}.
 More rigorously, we impose a general constraint:
 \begin{equation}
  \sup_{0<\epsilon<a} \int_0^L \frac{dx}{L}
    \abs[\bigg]{\frac{\varphi(x+\epsilon)-\varphi(x)}{\epsilon^q}}^2
  \le \frac{M^2}{L^{2q}},
  \label{HolderReg}
 \end{equation}
 where $a > 0$ is a constant that does not affect the estimation error in the limit of large $N$.
 The special case with $q=1$ reduces to Eq.~\eqref{DiffReg}.

 \paragraph{Estimation methods.}
 Given the target function $\varphi$ under the constraint~\eqref{HolderReg}, there exist estimation methods that ensure a finite estimation error $\delta$ defined in Eq.~\eqref{def-MSPE}.
 We here compare the following two different methods:

 Position-state (PS) method -- We estimate the individual phases $\varphi(x_j)$ at several positions $x_j$, and then computationally reconstruct the entire function.

 Wavenumber-state (WS) method -- We prepare a sufficiently large number of wavefunctions $\psi(x) \propto e^{i\varphi(x)}$, and estimate the function $\varphi$ by reconstructing the quantum state $\ket{\psi}$ by the quantum tomography.

 We find that the numbers of particles $N$ required for these two methods are the same up to a constant factor.
 
 \paragraph{Position-state method.}
 The PS method can be used when the target function is relatively small, say, $\abs{\varphi(x)}\le \pi/3$ for all $x$.
 In this case, we can circumvent the phase wrapping problem and employ a method analogous to the kernel density estimation~\cite{Stone1980}.

 In the first step, we sample $n_1$ positions $x_1, x_2, \dotsc, x_{n_1}$ in the interval $0\le x\le L$ with equal spacing. 
 Then, the phase $\varphi_j = \varphi(x_j)$ at each position $x_j$ is measured by using $n_2$ particles localized at $x_j$.
 The estimation error $\delta_\f1{ind}$ of the individual phase $\varphi_j$ is known~\cite{Kitaev1995,Higgins2007} as it is the quantum metrology on a scalar parameter; the SQL $\delta_\f1{ind} = O(n_2^{-1/2})$ is established by the probe $\frac{1}{\sqrt2}(\ket{x_j;-}+\ket{x_j;+})^{\otimes n_2}$ and the Heisenberg limit $\delta_\f1{ind} = O(n_2^{-1})$ by $\frac{1}{\sqrt2}(\ket{x_j;-}^{\otimes n_2}+\ket{x_j;+}^{\otimes n_2})$.
 Finally, the function estimator $\tilde{\varphi}(x)$ is computed from the individual estimators $\tilde{\varphi}_j$ by local linear smoothing~\cite{Wand1994}:
 \begin{equation}
  \tilde{\varphi}(x) = \sum_{j=1}^{n_1} \tilde{\varphi}_jf(x-x_j),
  \label{PS-lerp}
 \end{equation}
 where $f$ is a smoothing function.
 In the present case, we may just set $f(x) = 1$ for $x \le \frac{L}{2n_1}$ and $f(x)=0$ otherwise.
 This corresponds to the approximation by the value at the nearest site, i.e., we set $\tilde{\varphi}(x) = \tilde{\varphi}_j$ where $x_j$ is the point nearest to $x$.

 The estimation error can be decomposed into two parts: the statistical error $\delta_\f1{stat}$ caused by the measurement, and the deterministic error $\delta_\f1{det}$ due to smoothing.
 The balance between these errors can be tuned by the width $l$ of smoothing.
 The estimated value $\tilde{\varphi}_j(x)$ is of the same order as $\delta_\f1{ind}$: $\delta_\f1{stat} = O(n_2^{-1/2})$ for the SQL and $\delta_\f1{stat} = O(n_2^{-1})$ for the Heisenberg limit.
 On the other hand, the deterministic error $\delta_\f1{det}$ is the variation of $\varphi(x)$ within the width $\frac{L}{2n_1}$, which turns out to be
 \begin{math}
  \delta_\f1{det} = O(n_1^{-q}M)
 \end{math}
 by virtue of the constraint~\eqref{HolderReg}.
 
 For a given number of particles $N = n_1n_2$, the optimal accuracy is determined by the trade-off between $\delta_\f1{stat}$ and $\delta_\f1{det}$.
 As a consequence of Young's inequality, we obtain
 \begin{equation}
  \delta \ge O(n_1^{-q} M) + O(n_2^{-1/2})
   \ge O\bigl( (M^{1/q}N^{-1})^{q/(2q+1)} \bigr)
   \label{PS-SQL}
 \end{equation}
 for the SQL and
 \begin{equation}
  \delta \ge O(n_1^{-q} M) + O(n_2^{-1})
   \ge O\bigl( (M^{1/q}N^{-1})^{q/(q+1)} \bigr)
   \label{PS-HL}
 \end{equation}
 for the Heisenberg limit.
 Therefore, the overall estimation error $\delta$ is significantly larger than the traditional quantum limit, which is an expected feature of the function estimation.
 We note that entanglement of particles in different positions is not necessary to achieve the Heisenberg limit; such intersite entanglement does not enhance the estimation of linear parameters, as suggested in studies of quantum network sensors~\cite{Proctor2018,Ge2018}.
 \paragraph{Wavenumber-state method.}
 In the WS method, we begin with the wavenumber eigenstate with zero eigenvalue: 
 \begin{math}
  \int_0^L \frac{dx}{\sqrt{2L}}[\ket{x;-}^{\otimes n_\f1{p}} + \ket{x;+}^{\otimes n_\f1{p}}].
 \end{math}
 We use the one-particle state ($n_{\f1p}=1$) for the SQL and a multipartite EPR state ($n_{\f1p}>1$) for the Heisenberg limit.
 
 By the phase-shift gate $\hat{U}_\varphi$, one obtains the output probe state
 \begin{gather}
  \ket{S_\varphi} =
    \int_0^L \frac{dx}{\sqrt{2L}} \bigl[\ket{x;-}^{\otimes n_\f1{p}} + e^{in_\f1{p}\varphi(x)}\ket{x;+}^{\otimes n_\f1{p}}\bigr].
  \label{WS-output}
 \end{gather}
 The estimation is conducted by reconstructing $\ket{S_{\varphi}}$ as accurate as possible by measuring $n_{\f1c}$ copies of the probe state.
 For this purpose, we consider the projection $P_K$ onto the subspace of wavenumbers $k$ such that $\abs{k} \le 2\pi K/L$.
 Since the postselected state $\ket{S^*_\varphi}\propto P_K\ket{S_\varphi}$ belongs to a $(2K+1)$-dimensional Hilbert space, it can be identified by the quantum tomography.

 The error of the state reconstruction can be quantified by the infidelity
 \begin{math}
  1 - \abs{\braket{ S_{\varphi} }{ S_{\tilde\varphi} }},
 \end{math}
 where $\ket{S_{\tilde\varphi}}$ denotes the reconstructed state.
 In fact, we show in the Supplemental Material~%
 \footnote{See Supplemental Material for the rigorous derivation of 
 the biased Cram\'er-Rao bound, details of the explicit estimation procedures, and an application of Kitaev's method on function estimation.}
 that the MSPE has can be bounded by the expected infidelity as
 \begin{align}
  \delta^2 &\le \frac{\pi^2}{n_{\f1p}^{2}} \f3E[
    1 - \abs{\braket{ S_{\varphi} }{ S_{\tilde\varphi} }}
  ] \notag \\
  &\le \frac{\pi^2}{n_{\f1p}^{2}}(1 - \abs{\braket{ S_{\varphi} }{ S^*_{\varphi} }}^2) +
   \frac{\pi^2}{n_{\f1p}^{2}} \f3E[
    1 - \abs{\braket{ S_{\varphi}^* }{ S_{\tilde\varphi} }}^2
  ] \notag \\
  &= \delta^2_\f1{PS} + \delta^2_\f1{QT}.
  \label{MSPE-Infidelity}
 \end{align}%
 Here, the error is divided into the postselection part $\delta^2_\f1{PS}$ and the quantum-tomography part $\delta^2_\f1{QT}$.
 The postselection error can be bounded by the constraint~\eqref{HolderReg} as $\delta_\f1{PS} \le O(K^{-q}M)$~\cite{Quade1937,PROSSDORF1975,Note1},
 while the results of the finite-dimensional tomography imply $n_{\f1p}\delta_\f1{QT} \le O(K^{1/2} n_{\f1c}^{-1/2})$~\cite{Fujiwara1995}.

 When $n_\f1p = 1$ and $N = n_\f1c$, we have the trade-off relation between $\delta_\f1{PS}$ and $\delta_\f1{QT}$ for the SQL:
 \begin{equation}
  \delta_\f1{PS} = O(K^{-q}M),\quad \delta_\f1{QT} = O(K^{1/2}N^{-1/2}).
 \end{equation}
 By setting $n_1=K$ and $n_2 = N/K$, the errors
 $\delta_\f1{PS}$ and $\delta_\f1{QT}$ can be mapped to the errors
 $\delta_\f1{det}$ and $\delta_\f1{stat}$ in the PS method, respectively.
 Therefore, the SQL in the WS method reduces to that in Eq.~\eqref{PS-SQL} obtained by the PS method.

 The error bound can be lowered for $n_\f1p > 1$ and $N = n_\f1p n_\f1c$,
 while $K \le O(n_{\f1c})$ must be maintained in order to robustly conduct the quantum tomography.
 Therefore, the optimal trade-off relation for the Heisenberg limit is
 \begin{equation}
  \delta_\f1{PS} = O(n_\f1c^{-q}M),\quad \delta_\f1{QT} = O(n_\f1p^{-1}).
 \end{equation}
 By setting $n_1=n_{\f1c}$ and $n_2=n_{\f1p}$, this tradeoff relation corresponds exactly to the PS method, and we obtain the same Heisenberg limit as Eq.~\eqref{PS-HL}.
 However, there is a caveat that, in the output state~\eqref{WS-output} the phase ambiguity of $\varphi(x)$ modulo $2\pi/n_{\f1p}$ needs to be removed.
 We show in the Supplemental Material \cite{Note1} that this removal can be handled by analogy with the Kitaev's method~\cite{Kitaev1995}.
 
 \paragraph{Optimality of the SQL.}
 We have preposed the SQL~\eqref{PS-SQL} and the Heisenberg limits~\eqref{PS-HL} that can be achieved by both the PS and the WS methods.
 We show that these limits are in fact optimal; 
 any theoretical method is subject to the same bounds on the estimation error.

 We first derive the theoretical lower bound on the SQL.
 We consider the Fourier transform of the function $\varphi(x)$:
 \begin{equation}
  \varphi_k = \int_0^L \frac{dx}{L} e^{-2\pi ikx/L}\varphi(x).
 \end{equation}
 On the wavenumber basis, the constraint~\eqref{HolderReg} corresponds to the suppression of high-wavenumber components: $\varphi_k = o(k^{-q})$.
 In particular, the special constraint~\eqref{DiffReg} is equivalent to
 \begin{math}
  \sum_{k=1}^{\infty} k^2\abs{\varphi_{k}}^2 \le \frac{M^2}{8\pi^2},
 \end{math}
 which can be seen from Perseval's equality.
 A generalization of this argument leads to a sufficient condition on the constraint~\eqref{HolderReg}:
 \begin{equation}
  \sum_{k=1}^{\infty} k^{2q}\abs{\varphi_{k}}^2 \le \frac{M^2}{2c_0^2},
   \label{Fourier-asymptote}
 \end{equation}
 where $c_0 = 2\pi^q \max_{0\le x\le \pi} [x^{-q}\sin x]$ is a $q$-dependent constant.

 To utilize the known results in the discrete parameter estimation~\cite{Humphreys2013,Baumgratz2016}, we focus on the functions with only some low-wavenumber components.
 Using a $K$-dimensional vector $\f5u=(u_1,\dotsc,u_K)$, we parametrize the function $\varphi$ as
 \begin{equation}
  \varphi_{\f5u}(x) = \sum_{k=1}^{K} \sqrt{2}u_k\sin (2\pi ikx/L).
 \end{equation}
 Such function $\varphi_{\f5u}$ meets the constraint if $\nm{\f5u} \le \rho$ is satisfied for $\rho = c_0^{-1}MK^{-q}$.
 Since $\rho=o(1)$ and $\sqrt{2}\sin (2\pi ikx/L)$ forms an orthonormal basis, the MSPE $\delta^2$ of the function $\varphi$ can be bounded by MSE $\delta_{\f5u}^2$ of the vector $\f5u$. 

 Hence, instead of the function estimation, we may consider the vector estimation in which case the error can be evaluated by the quantum Cram\'{e}r-Rao bound (QCRB)~\cite{Holevo1973,Helstrom1969}.
 For an unbiased estimation, the QCRB is given as
 \begin{equation}
  \delta_{\f5u} \ge \delta_\f1{UUB} := K\bigl(\tr [J(\f5u)]\bigr)^{-1/2},
   \label{UBQCRB}
 \end{equation}
 where $[J(\f5u)]$ is the Fisher information matrix defined for the output probe state $\ket{\psi_{\f5u}}$ as
 \begin{equation}
  [J(\f5u)]_{jk} = 4\Re \bra[\big]{\tfrac{\partial}{\partial u_j} \psi_{\f5u}}
   \bigl[1-\ket{\psi_{\f5u}}\bra{\psi_{\f5u}}\bigr] \ket[\big]{\tfrac{\partial}{\partial u_k} \psi_{\f5u}}.
   \label{definition-QFI}
 \end{equation}
 Since the Fisher information is bounded from above by the SQL~\cite{Pang2014}:
 $[J(\f5u)]_{jj} \le 8N$ for each $1\le j\le K$, we obtain a uniform, unbiased bound
 \begin{math}
  \delta_\f1{UUB} = (K/8N)^{1/2}.
 \end{math}

 Although such \emph{uniform} bound is not applicable to the biased estimation,
 there exists the \emph{worst-case} biased bound $\delta_\f1{WBB}$~\cite{Note1}.
 Within the region $\nm{\f5u} \le \rho$, one can find a vector $\f5u$ satisfying
 \begin{equation}
  \delta_{\f5u} \ge \delta_\f1{WBB}, \quad
   \delta_\f1{WBB}^{-1} := \delta_\f1{UUB}^{-1} + \rho^{-1}.
  \label{QCRB}
 \end{equation}
 Since this biased version of QCRB holds for any integer $K\ge 1$, we choose $K$ that gives the maximal bound $\delta_\f1{WBB}$.
 This is satisfied when $\rho$ and $\delta_\f1{UUB}$ are comparable to each other --- this case holds when we set $K=O\bigl((M^2 N)^{1/(2q+1)}\bigr)$. 
 Hence the SQL is given as
 \begin{equation}
  \delta \ge c_1 (M^{1/q}N^{-1})^{q/(2q+1)},
   \label{SQL-lower}
 \end{equation}
 where the constant factor $c_1$ is lower-bounded by $\frac{q}{2(2q+1)^2} [c_0^{-1}q]^{2q/(2q+1)}$.
 \paragraph{Optimality of the Heisenberg limit.}
 We consider the case in which entanglement between at most ${n_\f1p}$ ($\ge 1$) particles is allowed.
 It is known that the quantum information of a probe state is maximal when their wavefunction is completely symmetric~\cite{Imai2007}.
 With completely symmetric probe states, the problem becomes equivalent to the estimation of an effective phase ${n_\f1p}\varphi(x)$ with $n_\f1c=n_\f1p^{-1}N$ separate particles;
 the probe states in~\eqref{WS-output} serve as an example for the WS method.


 Since the function of interest $\varphi$ is replaced by its effective one ${n_\f1p}\varphi$, the MSPE $\delta^2$ and the normalization constant $M$ are replaced by $n_\f1p^2\delta^2$ and ${n_\f1p} M$, respectively.  This argument leads to a generalized limit:
 \begin{align}
  n_\f1p\delta &\ge c_1 [({n_\f1p} M)^{1/q}(n_\f1p^{-1}N)^{-1}]^{q/(2q+1)} \notag \\
  &= c_1 (M n_\f1p^{q+1} N^{-q})^{1/(2q+1)}.
  \label{Reduced-SQL}
 \end{align}
 To restore the original function $\varphi(x)$ from the estimate of ${n_\f1p}\varphi(x)$, we need to resolve the phase ambiguity by $2\pi/{n_\f1p}$.
 For this purpose, the left-hand side of \eqref{Reduced-SQL} should not exceed $\pi$, giving
 \begin{equation}
  {n_\f1p} \le [(\pi c_1^{-1})^{(2q+1)/2}M^{-1}N^q]^{1/(q+1)}.
   \label{nu-upper-bound}
 \end{equation}
 With the maximal ${n_\f1p}$ substituted in \eqref{Reduced-SQL}, we obtain the Heisenberg limit
 \begin{equation}
  \delta \ge c_2 (M^{1/q}N^{-1})^{q/(q+1)},
  \label{HL-lower}
 \end{equation}
 where the constant $c_2$ is at least $(\pi^{-1}c_1)^{(2q+1)/(q+1)}$.
 \paragraph{Extension to smoother functions.}
 The degree of smoothness can further be extended into $q>1$, where the target function is known to be more than just differentiable.
 For an integer $m$ and $0<\sigma\le 1$ satisfying $q=m+\sigma$, the constraint for smoother functions is given as
 \begin{equation}
  \sup_{0<\epsilon<a} \int_0^L \frac{dx}{L}
    \abs[\bigg]{\frac{\varphi^{(m)}(x+\epsilon)-\varphi^{(m)}(x)}{\epsilon^{\sigma}}}^2
   \le \frac{M^2}{L^{2q}}.
   \label{GenHolderReg}
 \end{equation}

 Our results in the PS method and the optimality are also valid for $q>1$, thus leaving the quantum limits~\eqref{PS-SQL} and \eqref{PS-HL} unchanged.
 On the other hand, the straightforward extension of the WS method into $q>1$ does not work.
 Therefore, the asymptotic equivalence between the PS method and the WS method can be obtained only for $0<q\le 1$.
 See the Supplemental Material for more details~\cite{Note1}.

 \paragraph{Comparison with Gaussian signal estimation.}
 The error bounds we have obtained here are related to that of the Gaussian signal estimation~\cite{Berry2013,Berry2015,Dinani2017}, in which the time-dependent phase $\varphi_t$ is subject to a Gaussian process with the power spectrum $I(\omega)\sim\abs{\omega}^{-p}$.
 The estimation error of an instantaneous phase $\varphi_{t=0}$ is $O(\f2N^{-(p-1)/2p})$ for a coherent state and $O(\f2N^{-(p-1)/(p+1)})$ for a squeezed state, where $\f2N$ is the photon flux~\cite{Dinani2017}.
 This can exactly be mapped into the SQL $O(N^{-q/(2q+1)})$ and the Heisenberg limit $O(N^{-q/(q+1)})$ in our study by setting $p=2q+1$.
 In fact, almost all sample functions of the Gaussian process $\varphi_t$ satisfy Eq.~\eqref{GenHolderReg} by taking the large time span~\cite{Belayev1961,Kono1970}.
 Therefore, the estimation error of the Gaussian process is subject to the quantum metrology of function estimation.
 We suspect that this fact is related to the minmax theorem~\cite{Berger2013,Tanaka2012}, which explains the consistency between the Bayesian and non-Bayesian estimation methods, though no clear connection is established yet.
 
 \paragraph{Conclusion and outlook.}
 In this Letter, we have established the fundamental limits on function estimation subject to a bounded $q$th-order differentiability.
 The estimation error is bounded from below by $O( N^{-q/(2q+1)} )$ in the standard quantum limit and $O( N^{-q/(q+1)} )$ in the Heisenberg limit.
 These results can be reduced to the previous studies on the signal estimation in quantum optics~\cite{Berry2015,Dinani2017} when the target function is an infinitely extended stochastic process.
 We have presented two theoretical methods of the functional quantum metrology, both of which saturate the fundamental limits for $0<q\le 1$.
 
 This is a fundamental result for the effecient detection of functional structures --- continuous signals and images, for example --- which is a common target of estimation today.
 In fact, our results set theoretical bounds on various types of analysis relying on the function structure, such as model prediction or feature extraction~\cite{Moon2000,Prochazka2013}.
 On one hand, these bounds indicate the turning point where quantum methods outperform classical methods on functional data, with the scaling laws different from those obtained from parameter estimation.
 On the other hand, our result shows the optimal strategies for the quantum estimation of functions, such as an appropriate choice of temporal/spatial resolution or the size of entanglement.
 We note that choice of resolution is crucial in the real application~\cite{Tsang2016,Lupo2016,Tsang2017,Lu2018}, and what is more in the quantum case, we have seen that larger entanglement does not necessarily mean better accuracy.

 The framework presented here also enables further quantum information-theoretic analysis on functions, such as a the quantum version of the Nyquist--Shannon sampling theorem which concerns the exact equivalence between the position- and wavenumber-states in the signal detection, including the $O(1)$ prefactor that has remained undetermined. 
 \begin{acknowledgments}
 We gratefully acknowledge D.~W.~Berry for helpful discussions on the quantum signal estimation.  Special thanks are due to Y.~Ashida and R.~Hamazaki for critical advice. 
 This work was supported by a Grand-in-Aid for Scientific Research on innovative Areas ``Topological Materials Science'' (KAKENHI Grant No.~JP15H05855) from MEXT of Japan.
 N. K. was supported by the Leading Graduate Schools ``ALPS.''
 \end{acknowledgments} 
\bibliography{library}
\end{document}


\title{\maintitle}
\date{\today}
\author{Naoto Kura}\affiliation{\UTPhys}
\author{Masahito Ueda}\affiliation{\UTPhys}\affiliation{\RIKEN}\affiliation{\InsPI}
 \maketitle
 \section{Wavenumber-state method}
 We describe in detail the wavenumber-state method in the main text with a number of complementary arguments.
 First, we show the two inequalities in Eq.~(\ref*{!MSPE-Infidelity}) in the main text, in which the mean-square periodic error (MSPE) is decomposed into the postselection and quantum-tomography parts.
 We recall that the output state $\ket{S_\varphi} = \frac{1}{\sqrt2}(\ket{0}+\ket{\psi_\f1{out}})$ consists of the vacuum and the one-particle state $\psi_\f1{out}(x) = e^{i\varphi(x)}$.
 Hence we have
 \begin{align}
  \abs{ \braket{ S_{\varphi} }{ S_{\tilde\varphi} } }
  &= \abs[\bigg]{
    \int_0^L \frac{dx}{L}\frac{1+\exp \{i n_{\f1p}[\tilde{\varphi}(x)-\varphi(x)] \}}{2}
  } \notag \\
  &\le \int_0^L \frac{dx}{L} \cos \frac{n_{\f1p}}{2}[\tilde{\varphi}(x)-\varphi(x)]_{2\pi} \notag \\
  &\le \int_0^L \frac{dx}{L} \biggl( 1 - \frac{n_{\f1p}^2}{\pi^2}[\tilde{\varphi}(x)-\varphi(x)]_{2\pi}^2 \biggr)
   \le 1 - \frac{n_{\f1p}^2}{\pi^2}\delta^2,
 \end{align}
 which gives the first inequality
 \begin{math}
  \delta^2 \le \frac{\pi^2}{n_{\f1p}^{2}} \f3E[
    1 - \abs{\braket{ S_{\varphi} }{ S_{\tilde\varphi} }}
  ]. 
 \end{math}
 On the other hand, we can show 
 \begin{equation}
  \abs{ \braket{S_{\varphi}}{T} }^2 + \abs{ \braket{T}{S_{\tilde\varphi}} }^2
  \le 1 + \abs{ \braket{ S_{\varphi} }{ S_{\tilde\varphi} } }
 \end{equation}
 for an arbitrary state $\ket{T}$, whence the second inequality follows from
 \begin{equation}
  1 - \abs{ \braket{ S_{\varphi} }{ S_{\tilde\varphi} } } \le
  \bigl(1-\abs{ \braket{ S_{\varphi} }{ S^*_{\varphi} } }^2\bigr)
   + \bigl(1-\abs{ \braket{ S^*_{\varphi} }{ S_{\tilde\varphi} } }^2\bigr).
 \end{equation}

 We see that the postselection error
 $\delta^2_\f1{QT} = 1-\abs{ \braket{S_{\varphi}}{S^*_{\varphi}} }^2$ is at most $O(M^2/K^{2q})$.
 First, from $\ket{S^*_{\varphi}} \propto P_K\ket{S_\varphi}$, we obtain
 \begin{equation}
  \delta^2_\f1{QT} \le 1 - \bracket{S_\varphi}{P_K}{S_\varphi}.
 \end{equation}
 The right-hand side of this equation is equal to the squared amplitude of the high-wavenumber components of the wavefunction $\psi(x) = \frac{1}{\sqrt2}e^{i\varphi(x)}$.
 Since $\abs{e^{i\varphi_2}-e^{i\varphi_2}} \le \abs{\varphi_2 - \varphi_1}$ holds for any real $\varphi_1, \varphi_2$, the H\"older-class property of $\varphi$ is inherited by $\psi$:
 \begin{equation}
 \sup_{0<\epsilon<a}
 \int_0^L \frac{dx}{L} \abs[\bigg]{\frac{\psi(x+\epsilon)-\psi(x)}{\epsilon^q}}^2
  \le \frac{M^2}{2L^{2q}}.
  \label{HolderReg1}
 \end{equation}
 Therefore, the convergence of the Fourier series of $\psi$ can be evaluated according to Ref.~\cite{Quade1937}.
 \section{Phase estimation for the varying phase}
 In this section, we show that Kitaev's method~\cite{Kitaev1995,Higgins2007} can be extended to the varying phase, so that the wavenumber-state method can saturate the Heisenberg limit.
 As explained above, the Heisenberg limit can be saturated by using ${n_\f1p}$-body entanglement, with ${n_\f1p}=O\bigl( (M^{-1}N^{q})^{1/(q+1)} \bigr)$ indicating the maximal entanglement that can be exploited.

 For two phase-valued functions $\varphi$ and $\chi$, we define their distance $D(\chi, \varphi)$ by
 \begin{align}
  D^2( \chi, \varphi )
  = \int_0^L \frac{dx}{L} [\chi(x)-\varphi(x)]_{2\pi}^2.
 \end{align}
 Then, the MSPE can be written as $\delta = D(\tilde{\varphi},\varphi)$.
 Meanwhile, the obtained estimator $\tilde{\varphi}$ with this entanglement satisfies
 \begin{equation}
  D^2( {n_\f1p}\tilde{\varphi}, {n_\f1p}\varphi )
  = \int_0^L \frac{dx}{L} [{n_\f1p}\tilde{\varphi}(x)-{n_\f1p}\varphi(x)]_{2\pi}^2
   \le O( n_{\f1p}^2 (M^{1/q}N^{-1})^{2q/(q+1)} ).
   \label{180816_19Jun19}
 \end{equation}
 If we could replace $[{n_\f1p}\theta]_{2\pi}$ with ${n_\f1p} [\theta]_{2\pi}$ in Eq.~\eqref{180816_19Jun19},
 we would obtain the Heisenberg-limit MSPE
 \begin{math}
  \delta =  O\bigl( (M^{1/q}N^{-1})^{q/(q+1)} \bigr).
 \end{math}
 This replacement is not allowed, however, since $[{n_\f1p}\theta]_{2\pi}$ identifies those values of $\theta$ that differ by an integer multiple of $2\pi/{n_\f1p}$ rather than $2\pi$.
 Here, we show how to circumvent this problem.
 Throughout this section, we denote prefactors of $O(1)$ by $c_1$, $c_2$, etc.

 First, we recall the result for the standard quantum limit.
 Given $N$ separate particles, the deterministic and statical errors of estimation are
 \begin{align}
  \delta_\f1{det}  &= D(\varphi, \varphi^*) \le c_1 M\alpha^{q}, \\
  \delta_\f1{stat} &= D(\tilde{\varphi}, \varphi^*) \le c_2 (N\alpha)^{-1/2},
 \end{align}
 where $\alpha = l/L$ is the width of smoothing.
 If we fix $\alpha = c_3(MN)^{-1/(q+1)}$, the smoothed function $\varphi^*$ is also fixed, and the deterministic error becomes
 \begin{equation}
  \delta_\f1{det} = D(\varphi, \varphi^*) \le c_1c_3^{q} (M^{1/q}N^{-1})^{q/(q+1)}.
   \label{185553_16Aug18}
 \end{equation}

 Now, we consider $2^n$-body entanglement with integers $0\le n\le n_0$,
 where $2^{n_0} = c_4N\alpha = c_3c_4(M^{-1/q}N)^{q/(q+1)}$ corresponds to the maximal entanglement.
 By using $N_\f1{copy} = c_5^2\alpha^{-1}$ copies of a $2^n$-particle probe state, we obtain an estimator $\tilde{\varphi}_{(n)}$ satisfying
 \begin{equation}
  D( 2^n\tilde{\varphi}_{(n)}, 2^n\varphi^*)
   = c_2(N_{\f1{copy}}\alpha)^{-1/2} = c_2c_5^{-1} = O(1).
 \end{equation}

 Furthermore, if we compute the estimator $\tilde{\varphi}_{(n)}$
 repeatedly over $N_\f1{repeat} = c_6(n_0+1-n)$ times, with sufficiently large $c_5$ and $c_6$, we may apply the Chernoff bound so that the probability of the event
 \begin{math}
  [2^n\tilde{\varphi}_{(n)}(x) - 2^n\varphi^*(x) ]_{2\pi} > \pi/3
 \end{math}
 decreases exponentially.
 In particular, if we introduce the set
 \begin{equation}
  X_n = \{ 0\le x < L \mid [2^n\tilde{\varphi}_{(n)}(x) - 2^n\varphi^*(x) ]_{2\pi} \ge \pi/3 \},
 \end{equation}
 the expected value of the Lebesgue measure $\abs{X_n}$ is below $2^{-3(n_0-n)}L$.

 Finally, the estimated field $\tilde{\varphi}$ is determined from
 $\tilde{\varphi}_{(0)}$, $\tilde{\varphi}_{(1)}$, \ldots, $\tilde{\varphi}_{(n_0)}$
 in the following way.
 For every $x$, we set $\tilde{\varphi}(x)$ to be a phase $\theta$ satisfying
 \begin{equation}
  [2^n\tilde{\varphi}_{(n)}(x) - 2^n\theta ]_{2\pi} < \pi/3
   \label{182117_16Aug18}
 \end{equation}
 for $n=0,1,\ldots, m$.  Here, $m$ is the largest integer not exceeding $n_0$ such that the desired phase $\theta$ exists.

 The phase ambiguity up to $2\pi\cdot 2^{-m}$ is resolved by requiring \eqref{182117_16Aug18} for $0\le n\le m$.  This is always possible for $x$ that does not belong to any of $X_0,\ldots, X_{m}$.
 Therefore, the statistical error of $\tilde{\varphi}$ is evaluated as
 \begin{align}
  \delta_\f1{stat}^2 &= D^2(\tilde{\varphi},\tilde{\varphi}^*)
  \le \sum_{m=0}^{n_0-1} \frac{\abs{X_m}}{L}\biggl(\frac{2}{3}\pi\cdot 2^{-m}\biggr)^2
   + \biggl(\frac{2}{3}\pi\cdot 2^{-n_0}\biggr)^2 \notag \\
  &\le \sum_{m=0}^{n_0} \frac{4}{9}\pi^2\cdot 2^{-(3n_0-m)}
  \le \frac{8}{9}\pi^2 \cdot 2^{-2n_0} \le \frac{8}{9}\pi^2 (c_3c_4)^{-2}(M^{1/q}N^{-1})^{2q/(q+1)}. \label{185525_16Aug18}
 \end{align}
 The Heisenberg limit is obtained from Eqs.~\eqref{185553_16Aug18} and \eqref{185525_16Aug18}.

 Finally, we shows that at most $N$ particles are involved in this method.
 First, the number particles employed to compute the estimator $\tilde{\varphi}_{(m)}$ is
 \begin{equation}
  2^m N_\f1{copy} N_\f1{repeat} = c_5c_6\alpha^{-1}2^m(n_0+1-m).
   \label{135920_15Oct18}
 \end{equation}
 The summation of this term over $0\le m<n_0-1$ leads to $c_5c_6\alpha^{-1}(2^{n_0+1}-n_0-2) \le 2c_4c_5c_6 N$.
 Hence, the number of particles can be adjusted to be $N$ by tuning the parameter $c_4$.
 \section{Biased Cram\'er-Rao bound}
 In this section, we prove the biased Cram\'er-Rao bound given in (\ref*{!QCRB}) in the main text.
 Given a quantum state $\ket{\psi_{\f5u}}$ defined over $\f5u\in \f3R^K$, one is required to compute the estimator $\tilde{\f5u}$ such that the expected squared error
 \begin{equation}
  \delta^2_{\f5u} = \f3E\bigl[ \nm{\tilde{\f5u} - \f5u}^2 \mid \f5u \bigr]
 \end{equation}
 is small, where $\mathbb{E}[X\mid Y]$ is the expectation value of $X$ conditioned on the parameter $Y$.  In particular, we are interested in the worst-case error with respect to the target parameter $\f5u$ satisfying $\nm{\f5u}\le \rho$ with a prescribed radius $\rho$.
 Let $\f5u^* = \f3E[\tilde{\f5u}\mid \f5u]$ be the stochastic average of the estimator.
 When the estimation is unbiased, i.e.\@ $\f5u^* = \f5u$, the error bound is given by the Fisher information $J(\f5u)$:
 \begin{equation}
  \delta^2_{\f5u} \ge \tr [J^{-1}(\f5u)] \ge \frac{K^2}{\tr J(\f5u)},
 \end{equation}
 where the first inequality is the Cram\'er-Rao inequality and the second follows from Cauchy-Schwartz inequality.
 Thus we find a uniform, unbiased bound $\delta^2_\f1{UUB}$:
 \begin{equation}
  \delta^2_\f1{UUB} = \inf_{\nm{\f5u}\le \rho} \frac{K^2}{\tr J(\f5u)},
   \label{181255_7Jul18}
 \end{equation}
 and hence
 \begin{math}
  \delta^2_{\f5u} \ge \delta^2_\f1{UUB}
 \end{math}
 holds for any $\nm{\f5u}\le \rho$.

 In general, however, the estimator may be biased: $\f5u^*\ne \f5u$.
 In this case, the squared error can be decomposed into two terms:
 \begin{align}
  \delta^2_{\f5u} &= \nm{\f5u^*-\f5u}^2 + V(\f5u), \qquad
  V(\f5u) = \f3E\bigl[ \nm{\tilde{\f5u} - \f5u^*}^2 \bigr],
  \label{184424_7Jul18}
 \end{align}
 where the first term is the deterministic part called the \emph{bias} and the second term is the stochastic part called the \emph{variance}.
 The unbiased error bound can be applied only to the latter, giving
 \begin{equation}
  \tr V(\f5u) \ge \tr [D(\f5u)J^{-1}(\f5u)D^{\f1T}(\f5u)] \ge \frac{[\tr D(\f5u)]^2}{\tr J(\f5u)},
   \label{184525_7Jul18}
 \end{equation}
 where $[D(\f5u)]_{jk} = \partial \f5u^*_k / \partial \f5u_j$ is the Jacobian for the transformation $\f5u\mapsto \f5u^*$.
 Since the numerator of the lower bound depends on $\mathbf{u}$, it is possible that  the estimation error is lower than $\delta^2_\f1{UUB}$ given in \eqref{181255_7Jul18} at some $\f5u$ at the cost of augmented errors at another $\f5u$.

 Therefore, we need to handle the worst-case (maximum) error for a biased estimator, as opposed to the uniform (minimum) error for an unbiased estimator.
 In particular, we show the following theorem:
 \begin{thm}
  We define the worst-case biased bound $\delta^2_\f1{WBB}>0$ as
  \begin{equation}
   \frac{1}{\delta_\f1{WBB}} = \frac{1}{\delta_\f1{UUB}} + \frac{1}{\rho}.
  \end{equation}
  Then, $\delta^2_\f1{\f5u} \ge \delta^2_\f1{WBB}$ holds for at least one $\f5u$ such that $\nm{\f5u}\le \rho$.
 \end{thm}
 \begin{proof}
  Let us assume the contrary, and suppose that
  \begin{equation}
   \delta^2_\f1{\f5u} < \delta^2_\f1{WBB}
    \label{132426_15Oct18}
  \end{equation}
  holds for some $\nm{u}\le \rho$.
  By \eqref{181255_7Jul18}, \eqref{184424_7Jul18}, and \eqref{184525_7Jul18}, the estimation error is bounded as
  \begin{equation}
  \delta^2_{\f5u} \ge \nm{\f5u^*-\f5u}^2 +
   \biggl(\frac{\tr D(\f5u)}{K}\biggr)^2\delta^2_\f1{UUB}.
  \end{equation}
  Therefore, the inequality \eqref{132426_15Oct18} implies 
  \begin{align}
   b_\f1{max} &:= \sup_{\nm{\f5u}\le\rho} \nm{\f5u^*-\f5u}
   < \frac{\rho\delta_\f1{UBB}}{\delta_\f1{UBB} + \rho},
   \label{190410_7Jul18}\\
   D_\f1{max} &:= \sup_{\nm{\f5u}\le\rho} \frac{\tr D(\f5u)}{K}
   < \frac{\rho}{\delta_\f1{UBB} + \rho}.
   \label{185217_7Jul18}
  \end{align}

  Now, we consider the average of $\tr D(\f5u)/K$ over the $K$-dimensional ball $\nm{\f5u}\le \rho$ with volume $V\rho^K$.
  Since $\tr D(\f5u) = \nabla_{\f5u}\cdot \f5u^*$, we may employ the divergence theorem to obtain
  \begin{align}
   \int_{\nm{\f5u}\le \rho} \frac{d\f5u}{V\rho^K}\frac{\tr D(\f5u)}{K}
   &= \int_{\nm{\f5u} = \rho} \frac{d\f5S}{V\rho^K}\frac{\f5n\cdot \f5u^*}{K} \\
   &= \frac{1}{\rho}\int_{\nm{\f5u} = \rho} \frac{d\f5S}{KV\rho^{K-1}}\f5n\cdot \f5u^*.
  \end{align}
  Here $\f5n = \f5u/\rho$ is the normal vector of the $(K-1)$-dimensional sphere $\nm{\f5u}=\rho$ with area $mV\rho^{K-1}$.
  Moreover, we have $\tr D(\f5u)\le D_\f1{max}$ and $\f5n\cdot \f5u^* = \rho + \f5n\cdot(\f5u^*-\f5u) \ge \rho - b_\f1{max}$, which yield
  \begin{equation}
   D_\f1{max} \ge \frac{1}{\rho}(\rho - b_\f1{max}).
  \end{equation}
  On the other hand, combining inequalities \eqref{190410_7Jul18} and \eqref{185217_7Jul18} gives $D_\f1{max} < \frac{1}{\rho}(\rho - b_\f1{max})$.
  The theorem is proved by contradiction.
 \end{proof}

 \section{Detailed analysis of estimation error for \texorpdfstring{$q>1$}{q>1}}
 In this section, we conduct a rigorous analysis of the function estimation in the general classes of smoothness.
 \subsection{Generalization of the position-state method}
 For large $q$, the smoothing function $f(x)$ used in the position-state (PS) method must be altered, since we need to further suppress estimation error by using higher derivatives.
 Here, we describe the construction of the smoothing function for the generic case, with a rigorous proof on the estimation errors.
 
 We recall that the estimator of the phases $\varphi_j = \varphi(x_j)$ ($j=1,\dotsc,n_1$) is obtained from $n_2 = N/n_1$ particles.
 The estimated field can be written as a linear combination of $\tilde{\varphi}_{j}$,
 \begin{equation}
  \tilde{\varphi}(x) = \sum_{j=0}^{n_1-1}\tilde{\varphi}_jf(x-x_j),
   \label{164007_12Jul18}
 \end{equation}
 as written in the main text.
 Note, however, that the smoothing function $f(x)$ itself can be discontinuous.
 We rescale this function via
 \begin{math}
  f(x) = h(x/l),
 \end{math}
 where $l = \frac{(m+1)L}{2 n_1}$ sets the lengthscale of the smoothing.
 The integer $m$ is taken to satisfy $q = m+\sigma$ for some $0 < \sigma\le 1$.

 The function $h(y)$ is independent of the number $N$ of particles, and must fulfill the following three requirements so that
 the statistical error $\delta_\f1{stat}$ and
 the deterministic error $\delta_\f1{det}$ are both small:
 \begin{enumerate}
  \item $h(y)$ vanishes outside the interval $-1\le y<1$;
  \item $h$ is bounded as $\abs{h(y)} \le H$;
  \item For $k = 0,\dotsc,m$ and $0\le \theta<1$, the function satisfies
        \begin{equation}
         \sum_{j=0}^{m}
          \biggl[\frac{2(j+\theta)}{m+1}-1 \biggr]^k
          h\biggl( \frac{2(j+\theta)}{m+1}-1 \biggr) =
          \begin{cases}
           1 & (k=0); \\ 0 & (k=1,\dotsc,m).
          \end{cases}
          \label{Ragged-Orthogonal}
        \end{equation}
 \end{enumerate}
 Such a function exists.
 In fact, by the inverse function theorem and Vandelmonde's determinant formula, the linear equation system \eqref{Ragged-Orthogonal} can be solved for each $\theta$, where the solution can be taken to be continuous for $0\le \theta\le 1$ and thus bounded.

 Let us see that the statistical error and deterministic errors can both be bounded by using the above requirements.
 To begin with, the statistical error can be computed as
 \begin{equation}
  \delta^2_\f1{stat} = \sum_{j=1}^{n_1} \delta^2_{j,\f1{stat}}\int_0^L \frac{dx}{L}[f(x)]^2,
 \end{equation}
 where $\delta^2_{j,\f1{stat}}$ is the statistical error for the individual $\tilde{\varphi}_j$.
 The estimation error of a single phase scales as $\delta_{j,\f1{stat}} = O(n_2^{-X/2})$, where $X=1$ for the SQL and $X=2$ for the Heisenberg limit.
 Therefore, we obtain
 \begin{equation}
  \delta^2_\f1{stat} = \sum_{j=1}^{n_1} \delta^2_{j,\f1{stat}} \int_{-1}^{1} \frac{l dy}{L}\frac{[h(y)]^2}{(m+1)^2}
   = O(n_1n_2^{-X})\times \frac{H^2}{n_1 (m+1)} = O(n_2^{-X}).
   \label{171229_12Jul18}
 \end{equation}
 Therefore, the scaling law $\delta_\f1{stat} = O(n_2^{-X/2})$ in the main text is justified.

 Next, we examine the deterministic error.
 Within the statistical error computed above, the estimated function $\tilde{\varphi}(x)$ tends to give its estimated value:
 \begin{equation}
  \varphi^*(x) = \sum_{j=1}^{n_1}\varphi(x_j)f(x-x_j).
   \label{163407_28Sep18}
 \end{equation}
 The deterministic error is the MSPE of $\varphi^*$ to the target function $\varphi$.

 The Taylor expansion of $\varphi(x_j)$ around $x$ gives
 \begin{equation}
  \varphi(x_j) = \sum_{k=0}^{m} \frac{(x_j-x)^k}{k!}\varphi^{(k)}(x) +
   \int_{0}^{x_j-x} \frac{t^m dt}{(m-1)!} [\varphi^{(m)}(x+t)-\varphi^{(m)}(x)].
   \label{164651_21Jun19}
 \end{equation}
 However, Eq.~\eqref{Ragged-Orthogonal} implies that
 \begin{equation}
  \sum_{j=1}^{n_1} (x_j-x)^k f(x-x_j) =
   \begin{cases}
    1 & (k=0); \\ 0 & (k=1,\dotsc,m).
   \end{cases}
  \label{165545_29Sep18}
 \end{equation}
 
 Therefore, by substituting Eq.~\eqref{165545_29Sep18} in the summation~\eqref{164651_21Jun19}, all terms are canceled except for $k=0$ and the residual term:
 \begin{equation}
  \varphi^*(x) = \varphi(x) + \sum_{j=1}^{n_1}
   \int_{0}^{x_j-x} dt \frac{t^{m-1}}{(m-1)!} [\varphi^{(m)}(x+t)-\varphi^{(m)}(x)]
   h\biggl(\frac{x_j-x}{l}\biggr).
 \end{equation}
 Here, we consider the situation in which the offset of the sample points $x_1, \dotsc, x_{n_1}$ are taken at random.
 To be more specific, we pick $\alpha$ from the uniform distribution over $0\le \alpha < 1$ and set $x_j = (j+\alpha)\frac{L}{n_1}$.
 In this case, we can replace the average over $x_j$'s with the integration over $0\le x'\le L$, giving
 \begin{align}
  \varphi^*(x) - \varphi(x) &\cong n_1\int_{0}^{L}\frac{dx'}{L} h\biggl(\frac{x'-x}{l}\biggr) 
   \int_{0}^{x'-x} dt\frac{t^{m-1}}{(m-1)!} [\varphi^{(m)}(x+t)-\varphi^{(m)}(x)] \\
  &= n_1 \int_{-l}^l dt\frac{t^{m-1}}{(m-1)!} [\varphi^{(m)}(x+t)-\varphi^{(m)}(x)]
     \int_{x+l\sgn(t)}^{x+l\sgn(t)+t} \frac{dx'}{L} h\biggl(\frac{x'-x}{l}\biggr).
 \end{align}
 Here we have used the fact that the function $h$ vanishes for $\abs{x'-x} > l$.
 Hence
 \begin{align}
  \abs[\big]{\varphi^*(x) - \varphi(x)}^2 &\alt \biggl[
    n_1 \int_{-l}^l dt \frac{\abs{t}^{m-1}}{(m-1)!}
    \abs[\big]{\varphi^{(m)}(x+t)-\varphi^{(m)}(x)} \frac{l-\abs{t}}{L}H
  \biggr]^2 \\
  &\le \biggl(\frac{n_1 H}{L}\biggr)^2
  \int_{-l}^l dt \biggl[\frac{\abs{t}^{m-1} (l-\abs{t})}{(m-1)!}\biggr]^2
  \int_{-l}^l dt \bigl[ \varphi^{(m)}(x+t)-\varphi^{(m)}(x) \bigr]^2 \\
  &\le c_m H^2 l^{2m-1} \int_{-l}^l dt \bigl[ \varphi^{(m)}(x+t)-\varphi^{(m)}(x) \bigr]^2,
 \end{align}
 where $c_m = \frac{2m(m+1)^2}{2(m!)^2 (4m^2-1)}$.
 The deterministic error can be bounded as
 \begin{align}
  \delta^2_\f1{det} = &\int_0^L\frac{dx}{L} \abs{\varphi^*(x) - \varphi(x)}^2 \\
  &\alt c_m l^{2m-1}
  \int_{0}^{L} \frac{dx}{L} \int_{-l}^{l}dt [\varphi^{(m)}(x+t)-\varphi^{(m)}(x)]^2 \\
  &\le c_m l^{2m-1}
  \int_{-l}^{l}dt \abs{t}^{2\sigma}\frac{M^2}{L^{2m+2\sigma}}
  = c_m (l/L)^{2m+2\sigma} M^2 = c_m (\tfrac{m+1}{2})^{2q} n_1^{-2q} M^2.
 \end{align}
 Therefore we have shown $\delta_\f1{det} = O(n_1^{-q} M)$ for general $q$.
 \subsection{Difficulty in the wavenumber-state method}
 
 Unfortunately, the wavenumber-state method cannot be applied for $q>1$.
 This owes to the fact that the approximation $e^{i\varphi(x)} \approx 1 + i\varphi(x)$ does not hold beyond the first derivative in $x$.
 To be more concrete, we \emph{cannot} generalize the equation in \eqref{HolderReg1} for $m>0$:
 \begin{equation}
  \sup_{0<\epsilon<a} \int_0^L \frac{dx}{L}
   \abs[\bigg]{\frac{\psi^{(m)}(x+\epsilon)-\psi^{(m)}(x)}{\epsilon^\sigma}}^2
   \le \frac{M^2}{L^{2q}}.
   \label{HolderReg2}
 \end{equation}

 Nevertheless, we may achieve the SQL when the phase modulation is sufficiently small so that the approximation $e^{i\varphi(x)} \approx 1 + i\varphi(x)$ remains valid.
 To be specific, if we impose the condition
 \begin{equation}
  \int_0^L \frac{dx}{L} \abs[\big]{\varphi'(x)}^2 \le \frac{A^2}{L^2}
 \end{equation}
 and assume that the bound $A$ satisfies $A^{q}\ll M$, the SQL $\delta^2\sim (M^{1/q}N^{-1})^\frac{2q}{2q+1}$ can be saturated.
 In fact, the function $\varphi'(x)$ belongs to the Sobolev--Slobodeck{\ij} space $H^{q-1}\bigl([0,L]\bigr)$~\cite{Ziemer2012} and the modified version of Eq.~\eqref{HolderReg2}
 \begin{equation}
  \sup_{0<\epsilon<a} \int_0^L \frac{dx}{L}
    \abs[\bigg]{\frac{\psi^{(m)}(x+\epsilon)-\psi^{(m)}(x)}{\epsilon^\sigma}}^2
   \le \frac{c_1 A^{2q} + c_2M^{2}}{L^{2q}}
 \end{equation}
 can be derived.

 However, the same argument does not apply to the Heisenberg limit, where the estimation of ${n_\f1p}\varphi(x)$ is conducted by using ${n_\f1p}$-body entanglement.
 This is because the nonlinearity in the exponential function $e^{i{n_\f1p} \varphi(x)}$ becomes significant as the entanglement increases.
 In fact, the normalization parameters $A$ and $M$ are replaced by ${n_\f1p} A$ and ${n_\f1p} M$ for the ${n_\f1p}$-body entanglement, when $({n_\f1p} A)^{2q}$ becomes larger than ${n_\f1p} M$ for sufficiently large ${n_\f1p}$.

 \subsection{Asymptotic optimality of the error bounds}
 The proof of optimality in the main text only assumes that Eq.~(\ref*{!Fourier-asymptote}) in the main text:
 \begin{equation}
  \sum_{k=1}^{\infty} k^{2q}\abs{\varphi_{k}}^2 \le \frac{M^2}{2c_0^2},
   \label{Fourier-asymptote}
 \end{equation}
 is the sufficient condition for the target function in consideration.
 
 Therefore, we only need to show that Eq.~\eqref{Fourier-asymptote} with appropriate constant $c_0$ implies the generalized constraint (Eq.~(\ref*{!GenHolderReg}) in the main text):
 \begin{equation}
  \sup_{0<\epsilon<a} \int_0^L \frac{dx}{L}
    \abs[\bigg]{\frac{\varphi^{(m)}(x+\epsilon)-\varphi^{(m)}(x)}{\epsilon^{\sigma}}}^2
   \le \frac{M^2}{L^{2q}}.
   \label{GenHolderReg}
 \end{equation}
 Starting from the Fourier coefficient
 \begin{math}
  \varphi(x) = \sum_{k=-\infty}^{\infty} \varphi_k e^{2\pi ikx/L},
 \end{math}
  we obtain
 \begin{align}
  \frac{\varphi^{(m)}(x+\epsilon)-\varphi^{(m)}(x)}{\epsilon^\sigma}
  &= \sum_{k=-\infty}^{\infty} \frac{\varphi_k}{\epsilon^\sigma}
     \frac{\partial^m}{\partial x^m} [ e^{2\pi ik(x+\epsilon)/L}-e^{2\pi ikx/L} ] \notag \\
  &=  \frac{e^{2\pi ik\epsilon}-1}{\epsilon^\sigma} \biggl(\frac{2\pi ik}{L}\biggr)^m \varphi_k e^{2\pi ikx/L}.
 \end{align}
 Therefore, Perseval's equation implies
 \begin{align}
  \int_0^L \frac{dx}{L} \abs[\bigg]{
    \frac{\varphi^{(m)}(x+\epsilon)-\varphi^{(m)}(x)}{\epsilon^\sigma}
  }^2
  &= \sum_{k=-\infty}^{\infty}\abs[\bigg]{
       \frac{e^{2\pi ik\epsilon/L}-1}{\epsilon^\sigma}
       \biggl(\frac{2\pi ik}{L}\biggr)^m \varphi_k
     }^2 \notag \\
 & = 2\sum_{k=1}^{\infty} \abs[\bigg]{
       \frac{2\sin \pi k \epsilon/L}{\epsilon^\sigma}\biggl(\frac{2\pi k}{L}\biggr)^m
     }^2 \abs{\varphi_k}^2.
  \label{152121_22Jun19}
 \end{align}
 Here, we set $c_0 = 2(2\pi)^m\pi^\sigma \sup_{0\le x\le \pi} [x^{-\sigma} \sin x]$.
 Noting that
 \begin{equation}
  \abs[\bigg]{ \frac{\sin \pi k \epsilon}{\epsilon^\sigma} }
    = \biggl(\frac{\pi k}{L}\biggr)^\sigma \abs[\bigg]{ \frac{\sin (\pi k \epsilon/L)}{(\pi k \epsilon/L)^\sigma} }
    \le \biggl(\frac{\pi k}{L}\biggr)^\sigma \sup_{0\le x\le \pi} [x^{-\sigma} \sin x],
 \end{equation}
 the left-hand side of Eq.~\eqref{152121_22Jun19} can be estimated as
 \begin{align}
  \int_0^L \frac{dx}{L} \abs[\bigg]{
    \frac{\varphi^{(m)}(x+\epsilon)-\varphi^{(m)}(x)}{\epsilon^\sigma}
  }^2 \le 2 c_0^2 \sum_{k=1}^{\infty}\frac{k^{2q}}{L^{2q}} \abs{\varphi_k}^2.
 \end{align}
 Therefore, we have shown that \eqref{Fourier-asymptote} is a sufficient condition for Eq.~\eqref{GenHolderReg}.
 \bibliography{library}